\documentclass[10pt,pdftex]{scrartcl}

\usepackage{graphicx}
\usepackage{amsmath}
\usepackage{amssymb}

\begin{document}

\title{Numerical considerations about the SIR epidemic model with infection age}
\author{Ralph Brinks, Annika Hoyer\\Department of Statistics, Ludwig-Maximilians-University\\ Munich, Germany}
\date{}
\maketitle

\begin{abstract}
We analyse the infection-age-dependent SIR model from a numerical point of view. 
First, we present an algorithm for calculating the solution the 
infection-age-structured SIR model without demography 
of the background host. Second, we examine how and under 
which conditions, the conventional SIR model (without infection-age)
serves as a practical approximation to the infection-age SIR model. 
Special emphasis is given on the effective reproduction number.
\end{abstract}

\section*{Introduction}
We analyse the infection-age SIR model without demography 
of the background host, whose foundations date back to Kermack and McKendrick \cite{Ker27}. The focus is
primarily on numerical aspects. In the SIR model, the population is partitioned 
into three states, the \emph{susceptible} state, the \emph{infected} and the \emph{removed} 
state (the initial letters of the three states give the model's name `SIR'). 
The \emph{removed} state comprises people recovered and deceased from the \emph{infected} state. 
The numbers of the people in the susceptible and the removed states at time $t$ are denoted by 
$S(t)$ and $R(t)$, respectively. Furthermore, let $i(t, \tau)$ denote the density of 
infected people at time $t$ and duration $\tau$ since infection (i.e., the \emph{infection age}), 
such that the number $I(t)$ of infected at $t$ is
\begin{equation}\label{e:I}
I(t) = \int_0^\infty i(t, \tau) \mathrm{d}\tau.
\end{equation}

The transmission rate of the infected with infection age $\tau$ is $\beta(t, \tau)$
and the removal rate from the infectious stage is $\gamma(\tau).$ The rate $\gamma$
comprises mortality as well as remission.

We can formulate the following model equations for the infection-age SIR model \cite{Ina17}:

\begin{eqnarray}
   \frac{\mathrm{d} S(t)}{\mathrm{d} t} &=& -\lambda(t) \, S(t) \label{e:Sys1a} \\
   \left ( \frac{\partial}{\partial t} + \frac{\partial}{\partial \tau} \right ) \, i(t, \tau) &=& - \gamma(\tau) \, i(t, \tau) \label{e:Sys1b} \\
   \frac{\mathrm{d} R(t)}{\mathrm{d} t} &=& \int_0^\infty \gamma(\tau) \, i(t, \tau) \label{e:Sys1c}.
\end{eqnarray}

The incidence rate $\lambda$ in Eq. \eqref{e:Sys1a} is given by 
\begin{equation*}
\lambda(t) =  \int_0^\infty \beta(t, \tau) \, i(t, \tau) {\mathrm{d} \tau}
\end{equation*}
and is usually called \emph{force of infection} \cite{Ina17}.

System \eqref{e:Sys1a} -- \eqref{e:Sys1c} is accompanied with initial conditions
\begin{eqnarray}
S(0)       &=& S_0 \label{e:Sys2a}   \\
i(t, 0)    &=& \lambda(t) \, S(t) \\
i(0, \tau) &=& i_0(\tau) \\
i(0, 0)    &=& S_0 \, \int_0^\infty \beta(0, \tau) \, i_0(\tau) \mathrm{d} \tau \label{e:Sys2d}
\end{eqnarray}

with positive $S_0$ and integrable $i_0$. For later use, we additionally assume
that $i(t, \infty) := \lim_{\tau \rightarrow \infty} i(t, \tau) = 0.$ 
Condition \eqref{e:Sys2d} is called \emph{coupling equation} 
and guarantees that system \eqref{e:Sys1a} -- \eqref{e:Sys1c} is well-defined \cite{Che16}. Note that 
system \eqref{e:Sys1a} -- \eqref{e:Sys1c} is a generalisation of the SEIR model \cite[Section 5.5]{Ina17}. 

Detailed discussion of Equations \eqref{e:Sys1a} -- \eqref{e:Sys1c}
with initial conditions \eqref{e:Sys2a} -- \eqref{e:Sys2d} can be found 
in \cite[Chapter 5.3]{Ina17}. Using the definition 
\begin{equation}\label{e:DefvarGamma}
\varGamma(\tau) := \exp \left ( -\int_0^\tau \gamma(\sigma) \mathrm{d} \sigma \right ),
\end{equation}
the effective reproduction number $\mathcal{R}(t)$ is given by

\begin{equation}\label{e:Reff}
\mathcal{R}(t) =  S(t) \, \int_0^\infty \beta(t, \tau) \, \varGamma(\tau) \mathrm{d} \tau,
\end{equation}
\cite[Eq. (22),(23)]{Nis09}.

\bigskip

A typical situation is that the transmission rate $\beta(t, \tau)$ and the initial conditions
\eqref{e:Sys2a} -- \eqref{e:Sys2d} are given. Then, system \eqref{e:Sys1a} -- \eqref{e:Sys1c}
is solved and the effective reproduction number $\mathcal{R}$ is calculated by 
Eq. \eqref{e:Reff}. In a way, $\mathcal{R}$ can be seen as an indirect parameter for the 
infection-age-structured SIR model because it follows via Eq. \eqref{e:Reff} 
from the governing equations \eqref{e:Sys1a} -- \eqref{e:Sys1c} and \eqref{e:Sys2a} -- \eqref{e:Sys2d}. 
Sometimes, $\mathcal{R}(t)$ can be estimated more easily from population surveys than, for instance, 
the transmission rate $\beta(\tau, t)$. Then, the question arises if and how the 
infection-age-structured SIR model can be solved if the 
effective reproduction number $\mathcal{R}$ is given instead of $\beta$.

\bigskip

This article is organised as followed: 
First, we describe a numerical algorithm to solve the system given by 
Equations \eqref{e:Sys1a} -- \eqref{e:Sys1c} with initial conditions 
\eqref{e:Sys2a} -- \eqref{e:Sys2d} on a rectangular grid. Then, we consider an important 
special case where the transmission rate $\beta$ does not need to be known to solve 
system \eqref{e:Sys1a} -- \eqref{e:Sys1c}. Finally, we present an example to demonstrate
the theoretical considerations.

\section*{Numerical solution of the infection-age-structured SIR model}

\begin{figure}[ht]
  \centering
  \includegraphics[keepaspectratio,width=0.75\textwidth]{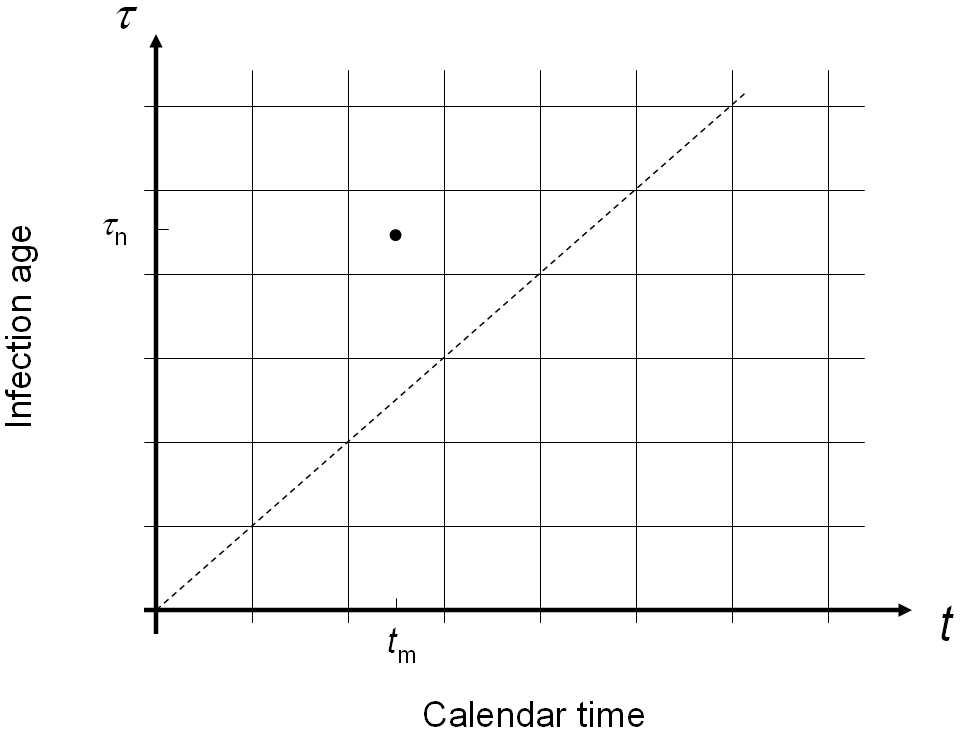}
\caption{Rectangular grid representing calendar time $t$ (abscissa) and infection-age 
$\tau$ (ordinate). The grid point $(t_m, \tau_n)$ above the main diagonal (dashed line) 
is highlighted.}
\label{fig:Grid}
\end{figure}

Assumed $i(t, \tau)$ has to be calculated on a rectangular grid 
$(t_m, \tau_n) = (m \times \delta_h, n \times \delta_h), ~m=0, \dots, M, ~n = 0, \dots, N,$
as depicted in Figure \ref{fig:Grid}. The grid points are assumed to be equidistant in $t$- 
and $\tau$-direction with distance $\delta_h > 0$.
A practical strategy for solving Equations \eqref{e:Sys1a} -- \eqref{e:Sys1c} 
with initial conditions \eqref{e:Sys2a} -- \eqref{e:Sys2d} is given by the following algorithm:
\begin{enumerate}
   \item Calculate $i(t_m, \tau_n) = i_0(\tau_n - t_m) \, \varGamma(\tau_n)$ for all $n \ge m$. 
   These are the incidence densities at the grid points located on and above the diagonal of the grid 
   (on and above the dashed line in Figure \ref{fig:Grid}).
   \item Given that $i(t_m, \tau_n)$ have been calculated on and above the diagonal, set $\ell := 0$
   and calculate $\lambda(t_{\ell+1})$ and $S(t_{\ell+1})$ to determine $i(t_{\ell+1}, 0)$.
   \item Calculate $i(t_{\ell+1+k}, \tau_{k}), ~k = 1, 2, \dots.$ The grid points $(t_{\ell+1+k}, \tau_{k})$ 
   are the points on a subdiagonal. We have $i(t_{\ell+1+k}, \tau_{k}) = i(t_{\ell+1}, 0) \, \varGamma(\tau_k)$.
   \item Set $\ell := \ell + 1$ and repeat steps 2 to 4 until the incidence density $i$
   has been calculated on all points $(t_m, \tau_n)~m=0, \dots, M, n = 0, \dots, N,$ on the grid.
\end{enumerate}

\bigskip

In case the transmission rate $\beta(t, \tau)$ only depends on calendar time $t,$ i.e., 
$\beta(t, \tau) = \beta(t),$ the force of infection can be written as
$$\lambda(t) = \beta(t) \, I(t) = \frac{\mathcal{R}(t) \, I(t)}{S(t) \, \int_0^\infty \, \varGamma(\tau) \mathrm{d} \tau}.$$
Then, System \eqref{e:Sys1a} -- \eqref{e:Sys1c} becomes explicitly dependent on $\mathcal{R}$. 
This means for given $\mathcal{R},$ 
the system can be solved, for instance by the algorithm above, such that the resulting
effective reproduction number equals the prescribed $\mathcal{R}.$ This is advantageous in situations, when 
the effective reproduction number $\mathcal{R}$ is known while the transmission rate $\beta$ is not.

\section*{Approximation of the age-structured SIR model by the conventional SIR model}

The situation described at the end of the last section, when the effective reproduction 
$\mathcal{R}(t)$ is known instead of the transmission rate $\beta(t, \tau)$, happens quite
frequently. Note that there are a variety of methods for estimating $\mathcal{R}$ from a 
time series of numbers of incident cases, see e.g. \cite{Cor13,Fra07}. The question arises, 
under which conditions system \eqref{e:Sys1a} -- \eqref{e:Sys1c} can be approximated by a 
simpler model that explicitly depends on $\mathcal{R}$.

A simpler model related to System \eqref{e:Sys1a} -- \eqref{e:Sys1c} is the conventional 
SIR model without demography of the background host:

\begin{eqnarray}
   \frac{\mathrm{d} S(t)}{\mathrm{d} t} &=& -\lambda(t) \, S(t) \label{e:Sys3a} \\
   \frac{\mathrm{d} I(t)}{\mathrm{d} t} &=& \lambda(t) \, S(t) - r(t) \, I(t) \label{e:Sys3b} \\
   \frac{\mathrm{d} R(t)}{\mathrm{d} t} &=& r(t) \, I(t) \label{e:Sys3c}
\end{eqnarray}

Using Leibniz' integral rule and Eq. \eqref{e:Sys2a}, the temporal derivative 
$\frac{\mathrm{d} I}{\mathrm{d} t}$ of the 
number of infected from Eq. \eqref{e:I} can be expressed as
\begin{eqnarray}
\frac{\mathrm{d} I(t)}{\mathrm{d} t} 
 &=& \frac{\mathrm{d}}{\mathrm{d} t} \int_0^\infty i(t, \tau) \mathrm{d}\tau 
                                       = \int_0^\infty \frac{\partial}{\partial t} i(t, \tau) \mathrm{d}\tau \nonumber \\
 &=& - \int_0^\infty \gamma(\tau) \, i(t, \tau) \mathrm{d}\tau - \int_0^\infty \frac{\partial}{\partial \tau} i(t, \tau) \mathrm{d}\tau \nonumber \\
 &=& - \int_0^\infty \gamma(\tau) \, i(t, \tau) \mathrm{d}\tau - i(t, \infty) + i(t, 0) \label{e:dI}.
\end{eqnarray}

As $i(t, \infty) = 0$ (see above) and $i(t, 0) = \lambda(t) \, S(t)$, Eq. \eqref{e:dI} reads as
\begin{equation}\label{e:I2}
\frac{\mathrm{d} I(t)}{\mathrm{d} t} = - \int_0^\infty \gamma(\tau) \, i(t, \tau) \mathrm{d}\tau + \lambda(t) \, S(t).
\end{equation}

It is reasonable to assume that the integral in Eq. \eqref{e:I2} has a finite upper bound $\omega < \infty$,  
because there are no infected people with infinite infection-age. As $i(t, \tau) \ge 0,$ the Mean Value 
Theorem for Definite Integrals \cite{Has17} guarantees existence of a 
$\tau^\star = \tau^\star(t) \in [0, \omega]$ such that
\begin{equation}\label{e:I3}
\frac{\mathrm{d} I(t)}{\mathrm{d} t} = - \gamma \bigl ( \tau^\star(t) \bigr ) \, I(t) + \lambda(t) \, S(t).
\end{equation}

So far, we could show that Eq. \eqref{e:Sys1b} from the age-structured SIR model 
can be approximated by Eq. \eqref{e:Sys3b} with $r(t) = \gamma (\tau^\star(t))$. If we can 
furthermore show that 
\begin{equation}\label{e:approx}
\lambda \, S = \mathcal{R} \, r \, I,
\end{equation}
we can reformulate Eq. \eqref{e:Sys3b} with an explicit dependency on $\mathcal{R}$. Assumed
Eq. \eqref{e:approx} holds true, we find
\begin{eqnarray}
   \frac{\mathrm{d} I(t)}{\mathrm{d} t} &=& - r(t) \, I(t) + \mathcal{R}(t) \, r(t) \, I(t) \nonumber \\
   &=& r(t) \, \bigl ( \mathcal{R}(t) - 1 \bigr ) \, I(t) \label{e:IReff}.
\end{eqnarray}

With the usual smoothness assumptions, Eq. \eqref{e:IReff} has the unique solution
\begin{equation}\label{e:IReff2}
I(t) = I(0) \, \exp \left (\int_0^t \Bigl [ r(\sigma) \, \bigl ( \mathcal{R}(\sigma) - 1 \bigl ) \Bigr ] \mathrm{d}\sigma \right ),
\end{equation}
where $I(0) = \int_0^\infty i_0(\tau) \mathrm{d}\tau$ (note that $i_0$ was assumed to be integrable).

Apart from their simplicity, Eqs. \eqref{e:IReff} and \eqref{e:IReff2} allow the common interpretation 
of the effective reproduction number $\mathcal{R}$: the number $I(t)$ of infected increases if and 
only if $\mathcal{R}(t) > 1.$

\bigskip

We have to examine the conditions such that Eq. \eqref{e:approx} holds true. As $i(t, \cdot)$ is non-negative, the Mean Value Theorem
for Definite Integrals applied to the left hand side of Eq. \eqref{e:approx} reads as
\begin{eqnarray}
   S(t) \, \lambda(t) &=&  S(t) \int_0^\infty \beta(t, \tau) \, i(t, \tau) \mathrm{d} \tau  \nonumber \\
                      &=&  S(t) \, \beta(t, \tau^\ast(t)) \, I(t) \label{e:approx2}
\end{eqnarray}
for  $\tau^\ast(t) \in [0, \omega].$ 

On the right hand side of Eq. \eqref{e:approx}, we have
\begin{equation}\label{e:approx3}
   \mathcal{R}(t) \, I(t) \, r(t) =  S(t) \, \beta(t, \tau'(t)) \, I(t) \, r(t) \, \int_0^{\omega'} \varGamma(\tau) \mathrm{d} \tau,
\end{equation}
where we assumed that $\beta(t, \cdot) \, \varGamma$ has a compact support $[0, \omega']$ and $\tau'(t) \in [0, \omega'].$

By comparing Eqs. \eqref{e:approx2} and \eqref{e:approx3}, we see that 
$r(t) \, \int_0^{\omega'} \varGamma(\tau) \mathrm{d} \tau = 1$ and $\beta(t, \tau^\ast(t)) = \beta(t, \tau'(t))$ 
implies $\lambda \, S = \mathcal{R} \, r \, I.$ Hence, for practical purposes if $r(t) = \gamma (\tau^\star(t))$ is close to 
$(\int_0^{\omega'} \varGamma(\tau) \mathrm{d} \tau)^{-1}$, we can expect that Eq. \eqref{e:IReff2} is a reasonable approximation for the 
number $I(t)$ of infected in the age-structured SIR model.

\section*{Example}
\begin{figure}[ht]
  \centering
  \includegraphics[keepaspectratio,width=0.9\textwidth]{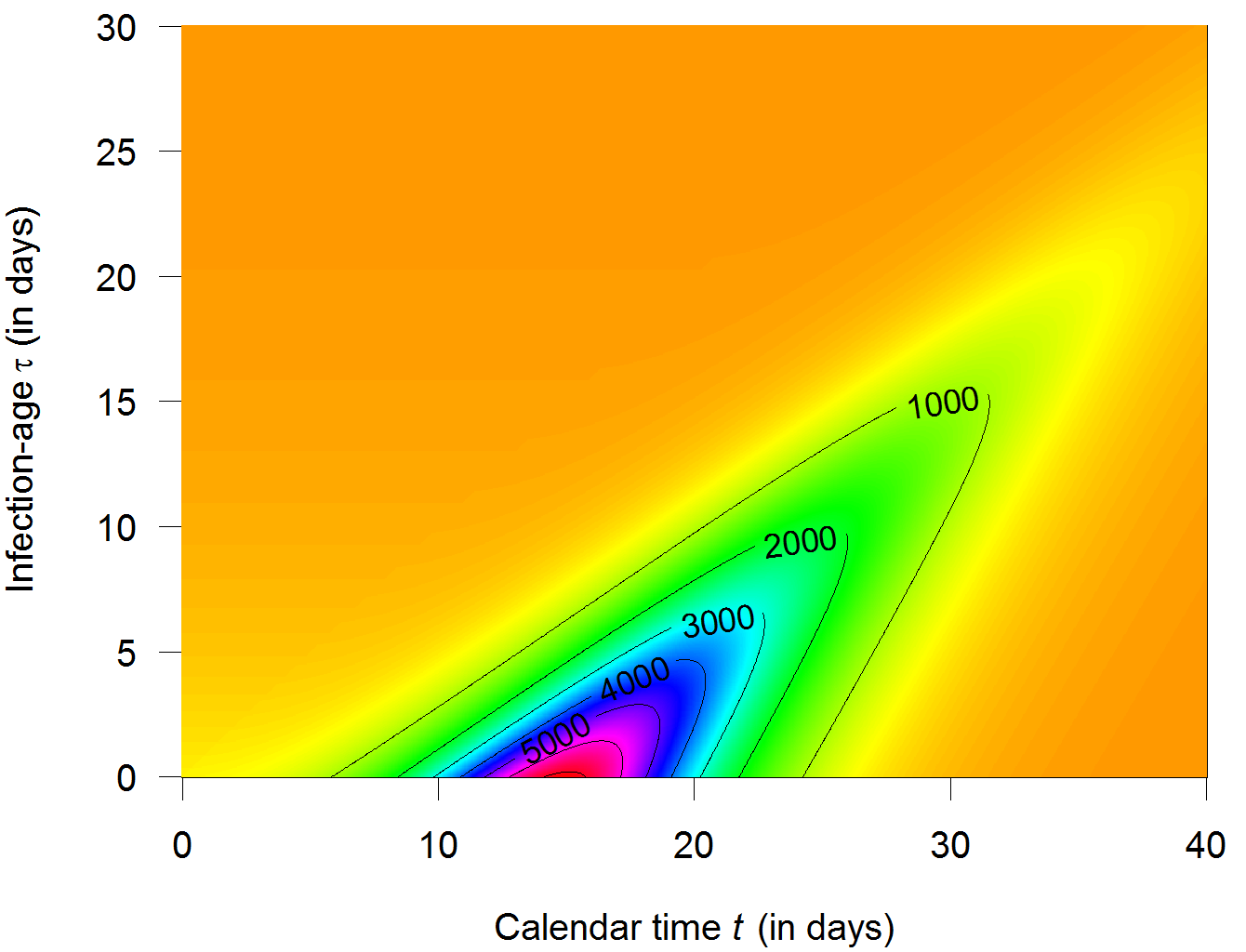}
\caption{Incidence-density $i(t, \tau)$ in the numerical example.}
\label{fig:Dens_i}
\end{figure}

We calculate the incidence-density $i$ on the grid $(t_m, \tau_n) = (m \times \delta_h, n \times \delta_h), 
~m = 0, \dots, M, ~n = 0, \dots, N,$ starting with $(t_0, \tau_0) = (0, 0)$, ending with 
$(t_M, \tau_N) = (40, 30)$ and equidistant stepsize $\delta_h = \tfrac{1}{24}$ (in units \emph{days}). 
The transmission rate $\beta(t, \tau)$ is assumed to be the product of two Gaussian functions:
$$\beta(t, \tau) = \nu_0 \times \Bigl ( \nu_1 + \nu_2 \, \exp \bigl ( -\bigl[(t-12)    /4  \bigr]^2/2 \bigr ) \Bigr ) \times 
                                \Bigl (                  \exp \bigl ( -\bigl[(\tau - 5)/1.5\bigr]^2/2 \bigr ) \Bigr ),$$
where $\nu_0 = 5\times 10^{-6}, \nu_1 = 1, \nu_2 = 3.$ The removal rate $\gamma$ is assumed to be constant 
$\gamma(\tau) = \tfrac{1}{8}$. The initial distribution $i_0$ is assumed to be $i_0(\tau) = 20 \times \varGamma(\tau),$ where
$\varGamma$ is defined in Eq. \eqref{e:DefvarGamma} with $\gamma(\tau) = \tfrac{1}{8}$. 
Figure \ref{fig:Dens_i} shows the resulting incidence-density $i(t, \tau)$
calculated with the Algorithm of the previous section.

\begin{figure}[ht]
  \centering
  \includegraphics[keepaspectratio,width=0.55\textwidth]{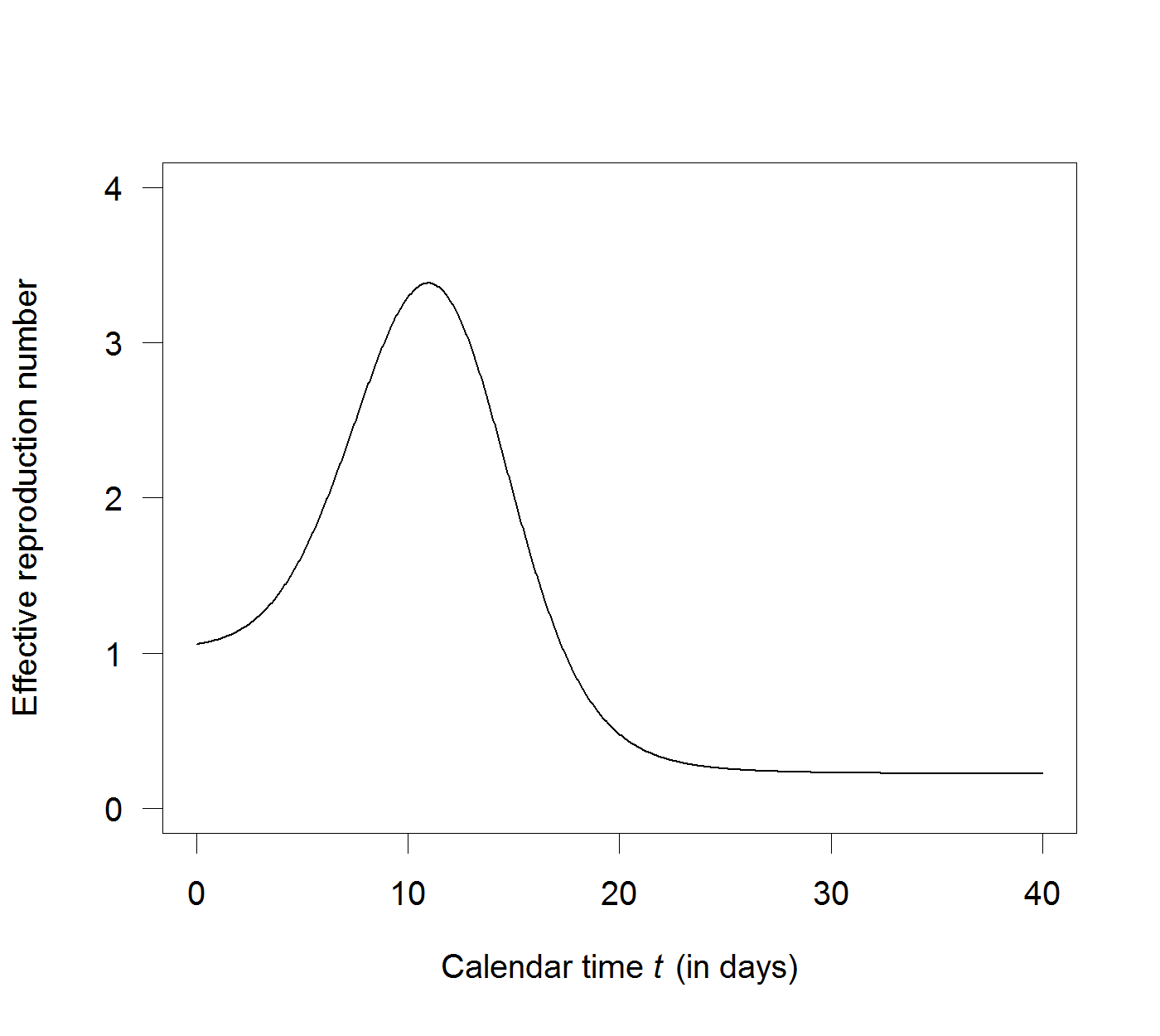}
\caption{Reproduction number $\mathcal{R}$ (ordinate) over calendar time $t$ 
in the example.}
\label{fig:Reff}
\end{figure}

\begin{figure}[ht]
  \centering
  \includegraphics[keepaspectratio,width=0.8\textwidth]{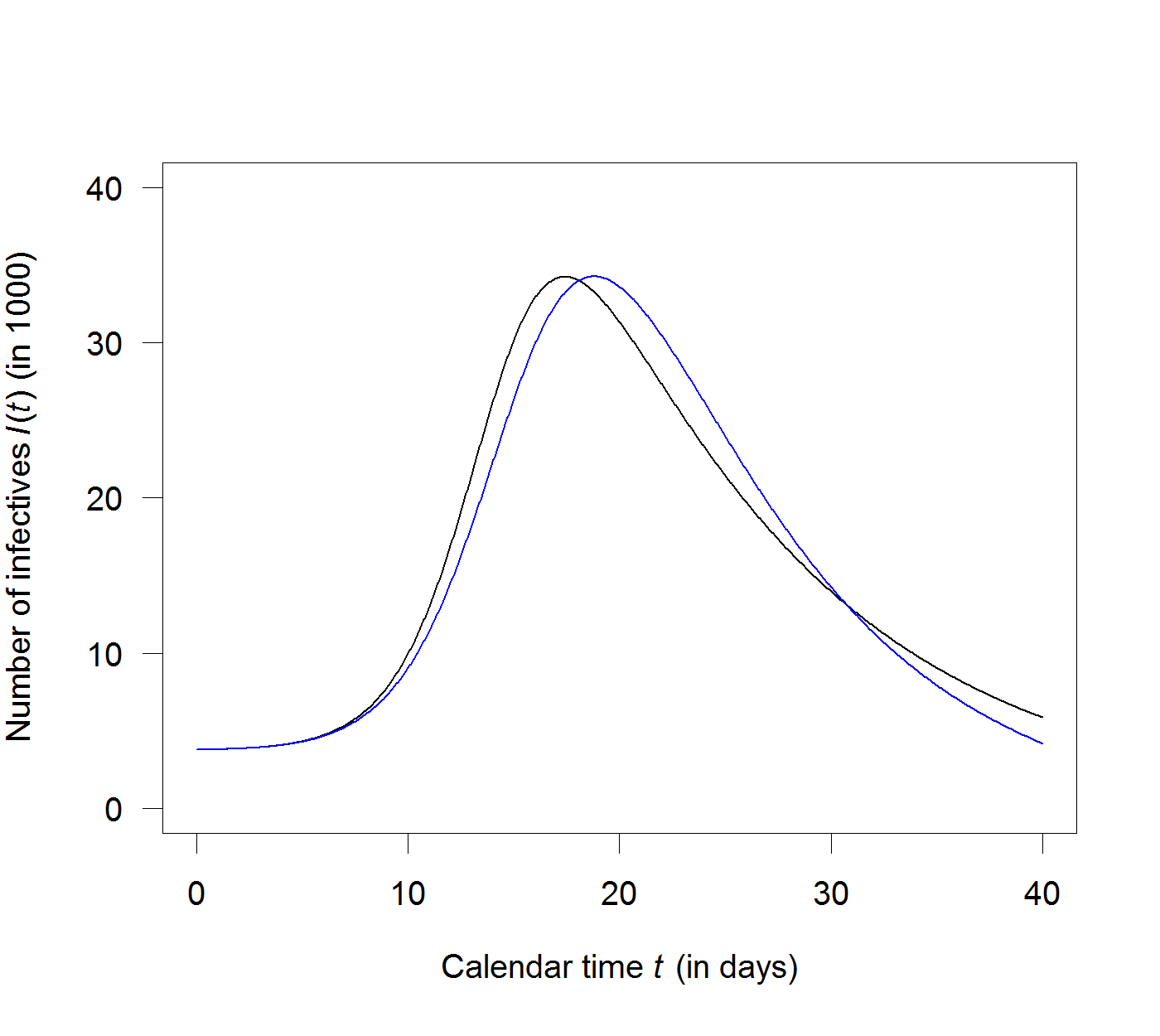}
\caption{Number $I$ of infectives (ordinate) over calendar time $t$ 
in the example. The blue curve corresponds to the exact solution 
(Eq. \eqref{e:I}) while the black curve is the approximation via 
Eq. \eqref{e:IReff2}.}
\label{fig:I}
\end{figure}

The resulting reproduction number $\mathcal{R}$ as calculated by Eq. \eqref{e:Reff}
is depicted in Figure \ref{fig:Reff}. If we try to approximate $I$ using $\mathcal{R}$ 
directly by Eq. \eqref{e:IReff2}, we obtain the graph as presented in Figure \ref{fig:I}.
The black curve corresponds to the approximated $I$ according to Eq. \eqref{e:IReff2} with
constant $r(t) = 0.1125$. For comparison, the exact $I$ calculated by \eqref{e:I}
is shown as blue curve. Although little effort has been spend to optimize the fit between
the exact and approximate $I$, the approximation is reasonably well.

\clearpage

\textbf{Contact:}\\
\verb"<firstname>.<lastname>@stat.uni-muenchen.de"\\
Ludwigstr. 33\\
D-80539 M\"unchen\\
Germany


{}


\begin{thebibliography}{}
\bibitem[Che16]{Che16}
Chen Y, Zou S, Yang J (2016) Global Analysis of an SIR Epidemic Model
with Infection Age and Saturated Incidence, Nonlin Ana: Real World App 30: 16-31.

\bibitem[Cor13]{Cor13}
Cori A, Ferguson NM, Fraser C, Cauchemez S (2013) A new framework 
and software to estimate time-varying reproduction numbers during epidemics. 
Amer J Epidem 178(9), 1505-1512.

\bibitem[Fra07]{Fra07} 
Fraser C (2007) Estimating Individual and Household Reproduction Numbers 
in an Emerging Epidemic. PLoS ONE 2(8): e758.

\bibitem[Has17]{Has17}
Hass JR,  Heil C, Weir MD (2017) Thomas' Calculus, 14th ed, Pearson, Harlow.

\bibitem[Ina17]{Ina17}
Inaba H (2017) Age-structured Population Dynamics in Demography
and Epidemiology, Springer, Singapore.

\bibitem[Ker27]{Ker27}
Kermack WO, McKendrick AG (1927) Contributions to the Mathematical
Theory of Epidemics, Proc Royal Soc 115A, 700-21.

\bibitem[Nis09]{Nis09}
Nishiura H, Chowell G (2009) The Effective Reproduction Number as a Prelude to 
Statistical Estimation of Time-Dependent Epidemic Trends, in: Chowell G, 
Hyman JM, Bettencourt LMA, Castillo-Chavez C: Mathmatical and Statistical
Estimation Approaches in Epidemiology, Springer, Dordrecht.

\end{thebibliography}
\end{document}